\documentclass[twocolumn,showpacs,preprintnumbers,amsmath,amssymb,floatfix]{revtex4-1}

\usepackage{graphicx}
\usepackage{dcolumn}
\usepackage{bm}

\renewcommand{\figurename}{Fig.}
\renewcommand{\tablename}{Table}

\def\SWTWOQFBP{\sin^2\theta_{\rm eff}^{\rm lept}}

\def\SM{{\rm SM}}

\def\be{\begin{eqnarray}
  }
\def\ee{\end{eqnarray}
  }

\begin{document}

%

\let\a=\alpha      \let\b=\beta       \let\c=\chi        \let\d=\delta
\let\e=\varepsilon \let\f=\varphi     \let\g=\gamma      \let\h=\eta
\let\k=\kappa      \let\l=\lambda     \let\m=\mu
\let\o=\omega      \let\r=\varrho     \let\s=\sigma
\let\t=\tau        \let\th=\vartheta  \let\y=\upsilon    \let\x=\xi
\let\z=\zeta       \let\io=\iota      \let\vp=\varpi     \let\ro=\rho
\let\ph=\phi       \let\ep=\epsilon   \let\te=\theta
\let\n=\nu
\let\D=\Delta   \let\F=\Phi    \let\G=\Gamma  \let\L=\Lambda
\let\P=\Pi     \let\Ps=\Psi   \let\Si=\Sigma
\let\Th=\Theta  \let\X=\Xi     \let\Y=\Upsilon

%

%

\def\cA{{\cal A}}                \def\cB{{\cal B}}
\def\cC{{\cal C}}                \def\cD{{\cal D}}
\def\cE{{\cal E}}                \def\cF{{\cal F}}
\def\cG{{\cal G}}                \def\cH{{\cal H}}
\def\cI{{\cal I}}                \def\cJ{{\cal J}}
\def\cK{{\cal K}}                \def\cL{{\cal L}}
\def\cM{{\cal M}}                \def\cN{{\cal N}}
\def\cO{{\cal O}}                \def\cP{{\cal P}}
\def\cQ{{\cal Q}}                \def\cR{{\cal R}}
\def\cS{{\cal S}}                \def\cT{{\cal T}}
\def\cU{{\cal U}}                \def\cV{{\cal V}}
\def\cW{{\cal W}}                \def\cX{{\cal X}}
\def\cY{{\cal Y}}                \def\cZ{{\cal Z}}
%

\newcommand{\Ns}{N\hspace{-4.7mm}\not\hspace{2.7mm}}
\newcommand{\qs}{q\hspace{-3.7mm}\not\hspace{3.4mm}}
\newcommand{\ps}{p\hspace{-3.3mm}\not\hspace{1.2mm}}
\newcommand{\ks}{k\hspace{-3.3mm}\not\hspace{1.2mm}}
\newcommand{\des}{\partial\hspace{-4.mm}\not\hspace{2.5mm}}
\newcommand{\desco}{D\hspace{-4mm}\not\hspace{2mm}}

\def\GF{G_F}
\def\mw{M_W}
\def\MBs{m_{B_s}}
\def\BBsfBstw{B_{B_s}f_{B_s}^2}
\def\lambdat{\lambda_t}
\def\Eta{\eta}
\def\Etap{\eta'}
\def\lambdatp{\lambda_{t'}}
\def\Szext{S_0(x_t)}
\def\Szextp{S_0(x_{t'})}
\def\Etatilde{\tilde\eta}
\def\Szetildextxtp{\tilde S_0(x_t,x_{t'})}
\def\*{\,}
\def\ie{{\it i.e. }}
\def\eg{{\it e.g. }}
\def\etal{{\it et al. }}
\def\no{\nonumber}
\def\ve{\varepsilon}
\newcommand{\NTU}{Department of Physics, National Taiwan University,
    Taipei, Taiwan 10617}
\newcommand{\NCTSn}{National Center for Theoretical Sciences, North Branch,
    National Taiwan University, Taipei, Taiwan 10617}


\title{\boldmath Flavor and CP Violation with Fourth Generations Revisited
}
\author{Wei-Shu Hou$^{1,2}$ and Chien-Yi Ma$^{1}$\\
{$^{1}$\NTU}\\
{$^{2}$\NCTSn}
 }
\date{\today}

\begin{abstract}
The Standard Model predicts a very small CP violation phase
$\sin2\Phi^{\rm SM}_{B_s} \simeq -0.04$
in $B_s$--$\bar B_s$ mixing. 
Any finite value of $\Phi_{B_s}$ measured
at the Tevatron would imply New Physics.
With recent hints for finite $\sin2\Phi_{B_s}$,
we reconsider the possibility of a 4th
generation. As recent direct search bounds have become considerably
heavier than 300 GeV, we take the $t'$ mass to be near the
unitarity bound of 500 GeV. Combining the measured values of
$\Delta m_{B_s}$ with ${\cal B}(B \to X_s\ell^+\ell^-)$, together
with typical $f_{B_s}$ values, we find a sizable $\sin2\Phi^{\rm
SM4}_{B_s} \sim -0.33$.
Using 
$m_{b'} = 480$ GeV, we extract the range
$0.06 < |V_{t'b}| < 0.13$ from the constraints of
$\Gamma(Z\to b\bar b)$, $\Delta m_{D}$ and
${\cal B}(K^+\to\pi^+\nu\bar\nu)$.
A future measurement of ${\cal B}(K_L\to\pi^0\nu\bar\nu)$ will determine $V_{t'd}$.
\end{abstract}

\maketitle

\section{Introduction}

There has been a recent mild revival~\cite{HHHMSU} for the 4
generation Standard Model (SM4). In good measure, this is due to
some hint~\cite{CDFDzero} for finite CP violation (CPV) phase
$\sin2\Phi_{B_s}$ at the Tevatron, which seems to resonate with
the unanticipated large deviation between direct CPV asymmetries,
observed by the B factories, between charged vs neutral $B$ meson
decays to $K\pi$ final states (the so-called $\Delta A_{K\pi}$
problem~\cite{DAKpi}). The 3 generation Standard Model (SM, or
SM3) predicts $\sin2\Phi^{\rm SM}_{B_s}
 \equiv \arg M_{12} \simeq \arg\,(V^*_{ts}V_{tb})^2
 \sim -\lambda^2\eta \simeq -0.04$,
where $\lambda$ and $\eta$ are parameters of the Wolfenstein
parametrization of the 3 generation CKM matrix~\cite{PDG}.
However, by its nondecoupling behavior, the heavy $t'$ quark is
especially suited to make impact on the above $b\to s$
processes~\cite{HLMN,HNS05,HNS07}.

Another reason of the mild revival is in regards electroweak
precision tests (EWPT). Some analyses show that even if the
oblique parameter $T$ is tuned to $0.232\pm 0.045$ in SM4, the
quality of the electroweak global fit still deteriorates
considerably  ($\Delta\chi^2=6.8$, disfavored at the $99\%$
CL)~\cite{PDG}.
However, the conclusion arises from the strong prejudice of
keeping $M_H$ fixed at the same SM3 value of 117 GeV. Several
papers~\cite{He,Kribs,Chanowitz2} demonstrate that, if $M_H$ is
taken as input variable, as is done for SM3, one could attain fits
that are sometimes better than SM3 in some parameter space.
Although this issue has recently been reopened~\cite{Erler}, as we
are concerned with the flavor and CP front, we will take the EWPT
issue just at that: an open question.

A third motivation for taking the 4th generation seriously is the
fundamental problem of CPV itself. While the unique CPV phase in
SM3 has been verified spectacularly by the B factories, but as
exemplified by the hint for $\sin2\Phi_{B_s}$, it may be just a
mirage. It is well known that the intrinsic CPV in SM3 falls short
of the requirement of the second Sakharov condition by a factor of
at least $10^{10}$. However, as noted by one of us, if one simply
extends SM3 to SM4, by being able to replace the rather light
second generation quark masses with the very heavy fourth
generation masses, the intrinsic CPV in SM4 may jump by
$10^{15}$~\cite{HouCJP} compared to SM3, and would seem sufficient
for generating the matter dominance of the Universe. Although the
third Sakharov condition remains an issue, this still elevates the
value for the pursuit of the 4th generation. The recent successful
collision of the Large Hadron Collider (LHC) at 7 TeV certainly
ups the ante of the search game, be it $\sin2\Phi_{B_s}$, or
direct search for the $t'$ and $b'$ quarks themselves.

Refs.~\cite{HNS07,HNS05} have studied flavor and CPV issues in
$B$, $K$ and $D$ systems. However, $m_{t'} = 300$ GeV was used,
qualified by the statement that a change in $m_{t'}$ would
correspond to some change in the CKM factors, with the gross
features retained. With the rising recent interest, and direct
search bounds now considerably above 300 GeV~\cite{CDFtp,CDFbp},
we revisit the flavor and CPV effects of a 4th generation with a
higher $t'$ mass. Our purpose is not to make a fit, since we deem
it premature, and could be misleading. Instead, we more or less
follow Refs.~\cite{HNS07} and~\cite{HNS05}, emphasizing salient
features. Also, although we touch upon the still developing
measurement of $D^0$--$\bar D^0$ mixing, we avoid incorporating
the uncontrolled long distance or hadronic effects such as $\Delta
A_{K\pi}$.

In the next section, we will discuss $\sin2\Phi_{B_s}$ by
comparing $\Delta m_{B_s}$ and ${\cal B}(b\to s\ell\ell)$, and
predict a possibly large deviation from SM3, due to heavy $t'$
interfering with $t$ through a nontrivial $V_{t's}^*V_{t'b}$. In
Sec.~III, we give an estimate of $V_{t'b}$, taking into
consideration $\G(Z\to b\bar b) /\G(Z\to{\rm hadrons})$, ${\cal
B}(K^+\to\pi^+\n\bar\n)$, $D^0$--$\bar D^0$ mixing and EWPT.
Taking a nominal value for $V_{t'b}$, a nominal value for
$V_{t's}$ is extracted, where critical dependence would be on
$m_{t'}$ and $f_{B_s}$. In Sec.~IV, adding the constraints of
$\ve_K$ and $\sin 2\Phi_{B_d}$, we discuss the correlations
between ${\cal B}(K_L\to\pi^0\n\bar\n)$ and $\sin 2\Phi_D$,
advocating the $K_L$ measurement as more critical in determining
$V_{t'd}$ in the future. We offer a brief conclusion in Sec.~V.

\section{Large \boldmath $\sin2\Phi_{B_s} ?$}

The measured CPV phase $\sin 2\Phi_{B_d}$
 ($\equiv\sin 2\phi_1 \equiv \sin2\b$)
via $B_d\to J/\psi K^0$ modes is consistent with SM, i.e. SM3.
However, recent measurements by the CDF and D\O\
experiments~\cite{CDFDzero} of the analogous $\sin 2\Phi_{B_s}$
 ($\equiv -\sin2\b_s \equiv \sin\phi_s$) in tagged $B_s^0 \to J/\psi\phi$
decays seem to give a large and negative value that is $2.1\, \s$
away from the SM expectation of $-0.04$. Though not yet
significant, the central value is tantalizingly close to a
prediction~\cite{HNS07} based on the 4th generation
interpretation~\cite{HLMN} of the observed $B^+$ vs $B^0\to K\pi$
direct CPV difference.

With four generations, the extra CKM product $V_{t's}^*V_{t'b}$
turns the familiar $b\to s$ unitarity triangle into a quadrangle:
\be
 V_{us}^*V_{ub}+V_{cs}^*V_{cb}+V_{ts}^*V_{tb}+V_{t's}^*V_{t'b} = 0.
 \label{sumutotp}
\ee
The $t'$ quark interferes with the top in the box diagram for
$B_s$--$\bar B_s$ mixing. We will use $\D m_{B_s}$, together with
the rare decay branching fraction ${\cal B}(b\to s\ell\ell)$,
which is dominated by the $Z$--penguin diagram, to constrain the
range of
\be
 \l_{t'} \equiv V_{t's}^*V_{t'b} \equiv r_{sb}e^{i\phi_{sb}},
 \label{ltp}
\ee
and gain a handle~\cite{HNS07,HNS05} on $\sin 2\Phi_{B_s}$.
Both the box and the $Z$--penguin diagrams are quite susceptible
to the nondecoupled $t'$ effects~\cite{HWS} through
$V_{t's}^*V_{t'b}$. The present study explores variations in
$f_{B_s}$ and $m_{t'}$.

Since the main source of information is from $B$ physics, we use
the convenient parametrization of Ref.~\cite{HSS87} for the
$4\times 4$ CKM matrix, where the 4th row and 3rd column are kept
particularly simple. We list the following elements for sake of
later discussions:
\be
V_{t'd} &=& -c_{24}c_{34}s_{14}e^{-i\phi_{db}},\label{vtpd}\\
V_{t's} &=& -c_{34}s_{24}\,e^{-i\phi_{sb}},\label{vtps}\\
V_{t'b} &=& -s_{34},\label{vtpb}\\
V_{t'b'} &=& c_{14}c_{24}c_{34},\label{vtpbp}\\
V_{ub'} &=& c_{12}c_{13}s_{14}\,e^{i\phi_{db}}+c_{13}c_{14}s_{12}s_{24}\,e^{i\phi_{sb}}\no\\
&&+c_{14}c_{24}s_{13}s_{34}\,e^{-i\phi_{ub}},\label{vubp}\\
V_{cb'} &=& c_{12}c_{14}c_{23}s_{24}\,e^{i\phi_{sb}}-c_{23}s_{12}s_{14}\,e^{i\phi_{db}}\no\\
&&+c_{13}c_{14}c_{24}s_{23}s_{34}-c_{14}s_{12}s_{13}s_{23}s_{24}\,e^{i(\phi_{sb}+\phi_{ub})}\no\\
&&-c_{12}s_{13}s_{14}s_{23}e^{i(\phi_{db}+\phi_{ub})}.\label{vcbp}
\ee
The form of $V_{tb'}$ is also more complicated, but
$V_{ub}=c_{34}s_{13}e^{-i\phi_{ub}}$, $V_{cb}=c_{13}c_{34}s_{23}$,
$V_{tb}=c_{13}c_{23}c_{34}$ are simple and close to the usual SM3
parametrization~\cite{PDG}. In the small angle limit, this allows
us to take the PDG values for $s_{12},\; s_{23},\; s_{13}$, as
well as $\phi_{ub}=\phi_3\cong 60^\circ$ as inputs, so
$V_{ij}\simeq V_{ij}^{\rm SM}$ for $i=u,c$ and $j=d,s,b$. From
(\ref{sumutotp}), one can also express
\be
 \l_t\equiv V_{ts}^*V_{tb}
     \simeq -r_{sb}e^{i\phi_{sb}} - \l_u^{\rm SM} - \l_c^{\rm SM}
 \label{lt}
\ee
in terms of $r_{sb}$ and $\phi_{sb}$. The notation of
$\phi_{sb}$, $\phi_{db}$ and $\phi_{ub}$ follows that of
Ref.~\cite{HNS05}.

The formula for $\D m_{B_s}$ is well known,
\be
 M_{12} &=& \frac{G_F^2M_W^2}{12\pi^2}m_{B_s}f_{B_s}^2\hat B_{B_s}
                  \Bigl[\l_t^2\eta S_0(x_t)  \no\\
   &&  + \eta'\l_{t'}^2S_0(x_{t'})
       + 2\tilde\eta\l_t\l_{t'}\tilde S_0(x_t,x_{t'})\Bigr].
 \label{M12Bs}
\ee
Let us first consider the case of $m_{t'} = 500$ GeV. Even though
$\D m_{B_s}^{\rm exp}=(17.77\pm0.12)\,{\rm ps}^{-1}$ is precisely
measured, the error for the current lattice value for $f_{B_s}\hat
B_{B_s}^{1/2}$ allows a large range for $r_{sb}$ and $\phi_{sb}$,
as shown in Fig.~1(a), where we have taken a recent result of
$f_{B_s}\hat B_{B_s}^{1/2} = 266(18)$ MeV~\cite{latHPQCD09} for
illustration. For $b\to s\ell\ell$ decay, we follow the NNLO
calculation of Ref.~\cite{Bobeth00}. However, as shown in
Fig.~1(b), here the experimental measurement of ${\cal B}^{\rm
exp}(b\to s\ell\ell)=(4.5\pm 1.0) \times 10^{-6}$~\cite{PDG} has a
sizable error, hence also allows a large range~\cite{Hiller} in
$r_{sb}$, $\phi_{sb}$.

\begin{figure}[t!]
\includegraphics[width=1.84in,angle=0]{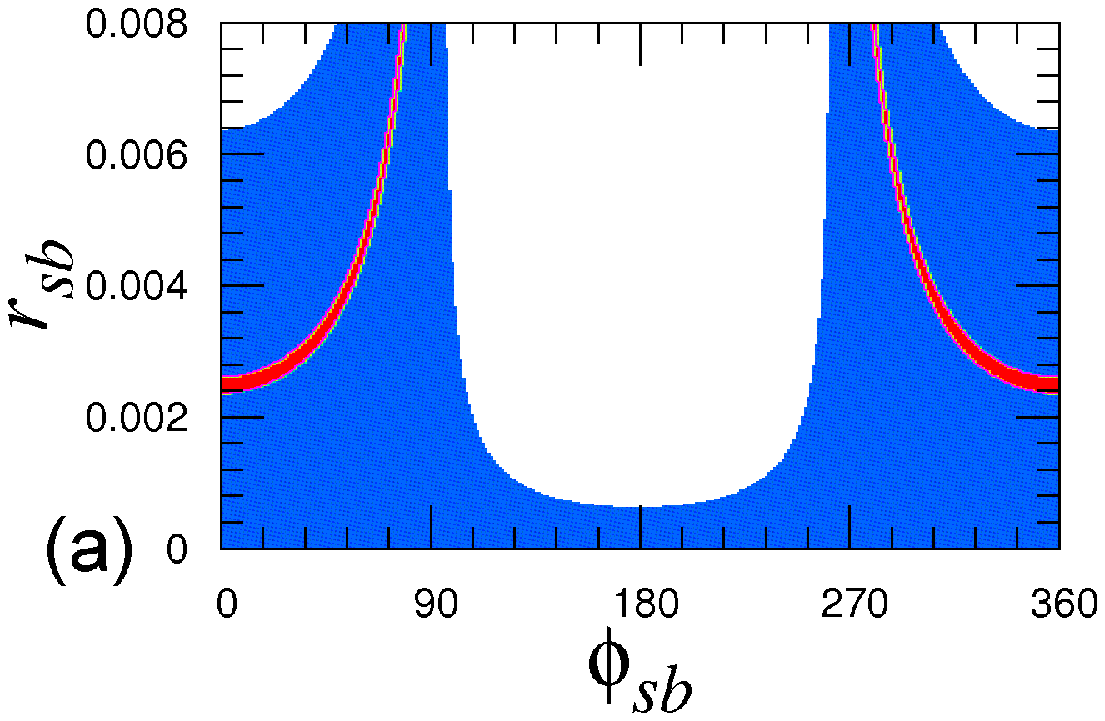}
\hspace{15mm}
\includegraphics[width=1.8in,angle=0]{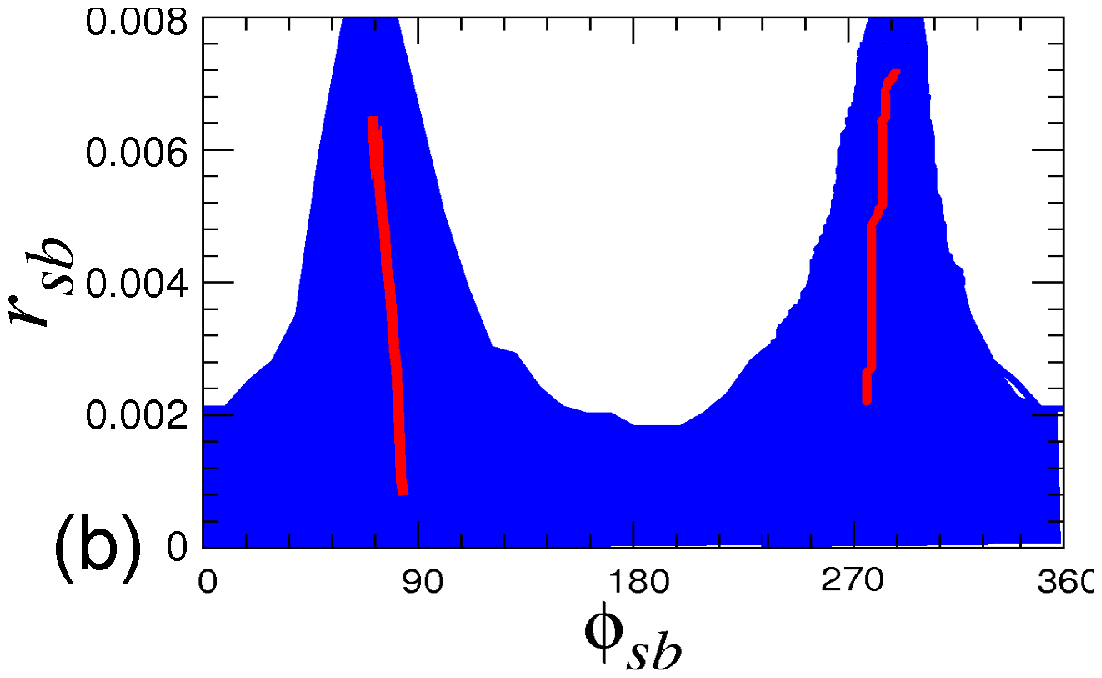}
\hspace{18mm}
\smallskip\smallskip\smallskip
\includegraphics[width=2.0in,angle=0]{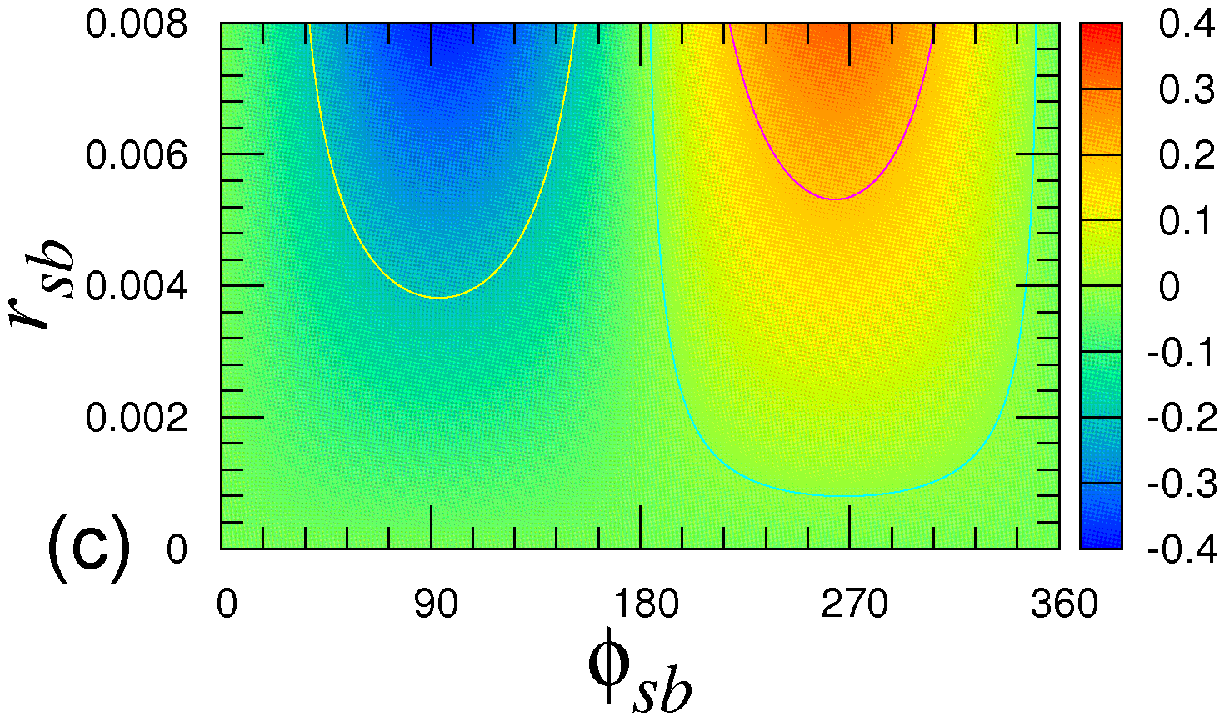}
 \caption{
 The allowed blue (or dark) range in $\phi_{sb}$--$r_{sb}$ from
  (a)~$\D m_{B_s}^{\rm exp}$ 
      due mainly to the lattice uncertainty in
      $f_{B_s}\hat B_{B_s}^{1/2} = 266(18)$ MeV, and
  (b) ${\cal B}^{\rm exp}(b\to s\ell\ell) = (4.5\pm1.0)\times10^{-6}$,
      where the red (or grey) lines correspond to taking
      the central values of 266 MeV and $4.5\times10^{-6}$, respectively.
 In (c), values of $\sin2\Phi_{B_s}$ are plotted over
  $\phi_{sb}$--$r_{sb}$ space.
 All plots are for $m_{t'}=500$ GeV.
 } \label{fig1}
\end{figure}

Comparing Figs.~1(a) with (b), and projecting onto
the $\sin2\Phi_{B_s}$ value plotted in Fig.~1(c), we still
have a lot of range for possible $\sin2\Phi_{B_s} \sim (-0.4,0.0)$
values. Note that positive $\sin2\Phi_{B_s}$, the righthand side
of Fig.~1(c), is ruled out by $\D {\cal
A}_{K\pi}$~\cite{HNS07,HNS05}.
For illustration, let us take the central values for $f_{B_s}\hat
B_{B_s}^{1/2}$ and ${\cal B}^{\rm exp}(b\to s\ell\ell)$,
illustrated by the (light) red lines in Figs.~1(a) and (b). We find
$\sin2\Phi_{B_s},\; r_{sb},\; \phi_{sb} = -0.33,\; 0.006,\;
75^\circ$, respectively. If we take the higher value of $f_{B_s}\hat
B_{B_s}^{1/2} = 295\,{\rm MeV}$, the same value as in the previous
study~\cite{HNS07}, we then have $\sin2\Phi_{B_s},\; r_{sb},\;
\phi_{sb} = -0.38,\; 0.010,\; 61^\circ$.

\begin{table}[t]
\begin{center}
\begin{tabular}{ccc}
  \hline\hline
  $V_{t's}^*V_{t'b}\backslash\sin2\Phi_{B_s}$
   & $\,f_{B_s}\hat B_{B_s}^{1/2}=266\,$MeV
   & $\,f_{B_s}\hat B_{B_s}^{1/2}=295\,$MeV \\
  \hline
  $m_{t'}=300$ GeV
   & $0.015\,e^{i81^\circ}\backslash-0.37$
   & $0.025\,e^{i70^\circ}\backslash-0.60$ \\
  \hline
  $m_{t'}=500$ GeV
   & $0.006\,e^{i75^\circ}\backslash-0.33$
   & $0.010\,e^{i61^\circ}\backslash-0.38$ \\
  \hline\hline
\end{tabular}
 \caption{
 Central values for $V_{t's}^*V_{t'b}$ and $\sin2\Phi_{B_s}$,
 corresponding to different $m_{t'}$ and $f_{B_s}\hat B_{B_s}^{1/2}$ values.}
\end{center}
\end{table}

The previous study was for $m_{t'} = 300$ GeV~\cite{HNS07,HNS05}.
Though seemingly ruled out by the Tevatron, this mass possibility
still needs to be crosschecked at the LHC. Following similar
procedures for this case, we find a larger $f_{B_s}\hat
B_{B_s}^{1/2}$ value would imply an even stronger $\sin2\Phi_{B_s}$.
Taking the central value of ${\cal B}^{\rm exp}(b\to s\ell\ell)$, we get
$\sin2\Phi_{B_s}$, $r_{sb}$, $\phi_{sb} = -0.37$, 0.015,
$81^\circ$ for the $f_{B_s}\hat B_{B_s}^{1/2} = 266$ MeV case,
compared with $-0.60,\; 0.025,\; 70^\circ$ for the $f_{B_s}\hat
B_{B_s}^{1/2} = 295$ MeV case (which roughly reproduces the result
of Ref.~\cite{HNS07}). Thus, if $-\sin2\Phi_{B_s}$ is found to be
larger than 0.5 or so, then larger $f_{B_s}\hat B_{B_s}^{1/2}$ values
would be preferred, and $t'$ mass would be likely closer to the
current Tevatron bounds.
On the other hand, the somewhat elaborate discussion here is in
the interest of predicting $\sin2\Phi_{B_s}$ when only $\Delta
m_{B_s}$ is known, which brings in a large uncertainty through
$f_{B_s}$. A future precision measurement would largely bypass the
$f_{B_s}$ dependence, and, together with knowledge of $m_{t'}$ and
improved measurement of ${\cal B}(b\to s\ell\ell)$,
should allow us good information on $V_{t's}^*V_{tb}$.

We summarize our results in Table I.
We note that for the 295 MeV case, the central value for $r_{sb}$
($\equiv|V_{t's}^*V_{tb}|)$ is considerably larger than for the
266 MeV case. This is because the SM3 value for $\D m_{B_s}$ is
already much higher than the experimental value, hence one would
need a larger $t'$ effect to compensate and bring it down. Higher
$r_{sb}$, however, will raise the lower bound of $|V_{t'b}|$,
which we now turn to discuss.

\section{Upper and lower bounds on \boldmath $|V_{t'b}|$}

An upper bound on $V_{t'b}$ comes from $R_b = \Gamma(Z\to b\bar
b)/\Gamma(Z\to{\rm hadrons})$ due to the loop diagram with $t'$.
Following Ref.~\cite{Yanir},
we find
%
\be
 \vert V_{tb}\vert^2 + 3.4 \vert V_{t^\prime b}\vert^2 < 1.14,
  && \ \ \ \
 m_{t^\prime} = 300 \;{\rm GeV}, \\
%
 \vert V_{tb}\vert^2 + 9.6 \vert V_{t^\prime b}\vert^2 < 1.14,
  && \ \ \ \
 m_{t^\prime} = 500 \;{\rm GeV}.
 \label{eq:Zbb}
\ee 
Applying the relatively good approximation
$|V_{tb}|^2\simeq1-|V_{t'b}|^2$, we get
%
\be
 |V_{t'b}|\le 0.13\; (0.24),\ \ \ \ \ m_{t^\prime} = 500\; (300) \;{\rm GeV}.
\ee
These upper bounds are given in Table~II. Note, of course,
that these bounds do not depend on $f_{B_s}$.

A lower bound on $\vert V_{t'b}\vert$ can arise from considering
${\cal B}(K^+\to\pi^+\nu\bar\nu)$ and $D^0$--$\bar D^0$ mixing
together. For the former, we use~\cite{BSU},
\be
 && \kappa_+|V_{us}|^{-10}
    \left|\lambda_c^{ds}|V_{us}|^4P_c
       +  \lambda_t^{ds}\eta_t X_0(x_t)
       +  \lambda_{t^{\prime}}^{ds}
          \eta_{t^{\prime}} X_0(x_{t^{\prime}}) \right|^2\no\\
 && \hspace{3cm}<3.6\times10^{-10}\;({\rm 90\%\,\,CL})
 \label{eq:kppipnn},
\ee
with $\lambda_{q}^{ds} \equiv V_{qd}V_{qs}^{*}$, and the 90\% CL
bound is from ${\cal B}^{\rm exp}(K^+\to\pi^+\nu\bar\nu) =
(1.73^{+1.15}_{-1.05}) \times 10^{-10}$~\cite{Artamonov}.
We define $V_{t'd}^{*}V_{t's} \equiv r_{ds}e^{i\phi_{ds}}$.

\begin{table}[t!]
\begin{center}
\begin{tabular}{ccc}
  \hline\hline
  bound on $|V_{t'b}|\ \ $ & $\ f_{B_s}\hat B_{B_s}^{1/2} = 266$ MeV
                           & $\ f_{B_s}\hat B_{B_s}^{1/2} = 295$ MeV \\
  \hline
  $m_{t'}=300$ GeV & (0.12, 0.24) & (0.20, 0.24) \\
  \hline
  $m_{t'}=500$ GeV & (0.06, 0.13) & (0.10, 0.13) \\
  \hline\hline
\end{tabular}
 \caption{Bounds on $|V_{t'b}|$ for different
 $m_{t'}$ (with $m_{b'}$ taken 20 GeV lower) and
 $f_{B_s}\hat B_{B_s}^{1/2}$ values.
 Note that the upper bound arising from $Z\to b\bar b$
 does not depend on $f_{B_s}\hat B_{B_s}^{1/2}$.
 The lower bounds arise from a possible tension between
 ${\cal B}(K^+\to\pi^+\nu\bar\nu)$ and $D^0$--$\bar D^0$ mixing.
 See text for discussion.}
\end{center}
\end{table}

For $\Delta m_D$, where $b'$ enters the loop, we follow the
formulas and ansatz in Ref.~\cite{Bobrowski,Petrov},
\be
 M_{12}^{D} &\propto&
  \lambda_s^2S_0(x_s) + 2\lambda_s\lambda_bS(x_s,x_b) + \lambda_b^2 S_0(x_b) + LD \no\\
 && + 2\lambda_s\lambda_{b'}S(x_s,x_{b'}) + 2\lambda_b\lambda_{b'}S(x_b,x_{b'}) + LD \no\\
 && + \lambda_{b'}^2S_0(x_{b'}),
 \label{m12d0}
\ee
%
where here $\lambda_{q} \equiv V_{uq}^{*}V_{cq}$. The first three
terms of the first line are the short distance SM3 contributions.
But experiment suggest sizable long distance (LD) contributions,
since $y_D$ is comparable~\cite{HFAG08} to $x_D$. Indeed, current
data is consistent with $D^0$--$\bar D^0$ mixing as due entirely
to LD effect. The second line involves both 4th and a lower
generation appearing in the box, but even here, there could be LD
effects. To allow for these two types of LD effects, we take the
purely short distance $|V_{ub'}^*V_{cb'}|^2S_0(x_{b'})$, i.e. the
last term, and equate it with $x_D^{\rm exp}$, but enlarging it by
a factor of 3. We then find
%
\be
 \left|V_{ub'}^*V_{cb'}\right| < (3.45^{-0.27}_{+0.35}) \times 10^{-3},
 \label{dda}
\ee 
for $m_{b'} = 260 \pm 30$ GeV, and
%
\be
 \left|V_{ub'}^*V_{cb'}\right| < (2.20^{-0.12}_{+0.13}) \times 10^{-3},
 \label{ddb}
\ee 
for $m_{b'} = 460 \pm 30$ GeV, where we have applied the latest
experimental value of $x_D^{\rm exp} = (9.1^{+2.5}_{-2.6}) \times
10^{-3}$~\cite{HFAG08}. The range for $m_{b'}$ contains the
sample value we would use for illustration.

With these set up, we can now discuss how a lower bound on
$|V_{t'b}|$ could arise.
From Eqs.~(\ref{vtpd})--(\ref{vcbp}), $V_{t'b},\; V_{t's},\;
V_{t'd}$ are proportional to $s_{34},\; s_{24},\; s_{14}$,
respectively, and $|V_{cb'}|\simeq|V_{t's}|$ if $s_{24}$ is not
unduly small. But it is less likely that $V_{ub'}\propto s_{14}$
would hold, since the likely larger angles $s_{24}$ and $s_{34}$
enter modulated only by factors of $s_{12}$ and $s_{13}$,
respectively, where $s_{12} \cong \lambda \simeq 0.2$ is not
particularly small. So, if $|V_{t's}^*V_{t'b}|$ is held {\it
fixed} (in the context of definite $m_{t'}$ and
$\sin2\Phi_{B_s}$), as $|V_{t'b}|\simeq s_{34}$ is lowered,
$|V_{t's}|\simeq s_{24}$ would grow. To satisfy the constraint of
Eq. (\ref{eq:kppipnn}), one would have to reduce $|V_{t'd}|\simeq
s_{14}$. But then, from Eq. (\ref{vubp}), $|V_{ub'}|$ would likely
rise and cause tension with Eqs.~(\ref{dda}) and (\ref{ddb}).
The form of Eq.~(\ref{vubp}), which is from the parametrization of
Ref.~\cite{HSS87}, helps in elucidating this effect. With $s_{14}$
constrained small while $s_{24}$ looming larger, the
$s_{12}s_{24}$ term would likely dominate $|V_{ub'}|$ (remember,
$s_{34}$ is pushed lower, and it is further modulated by $s_{13}$
which is the strength of $|V_{ub}| \simeq 0.003$), while
$|V_{cb'}| \simeq |V_{t's}|\simeq s_{24}$, hence the $\Delta m_D$
constraint of Eq.~(\ref{ddb}) becomes hard to satisfy.

\begin{figure}[t!]
\begin{center}
\includegraphics[width=2.0in,angle=0]{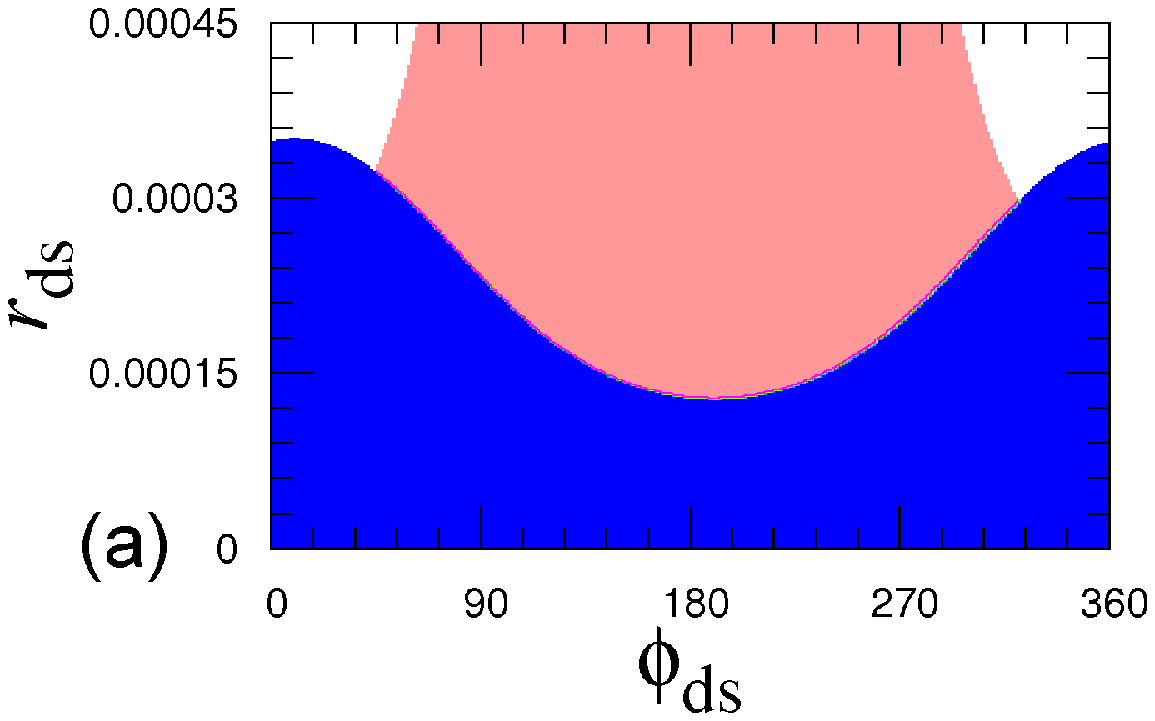}
\hspace{5mm}
\includegraphics[width=2.0in,angle=0]{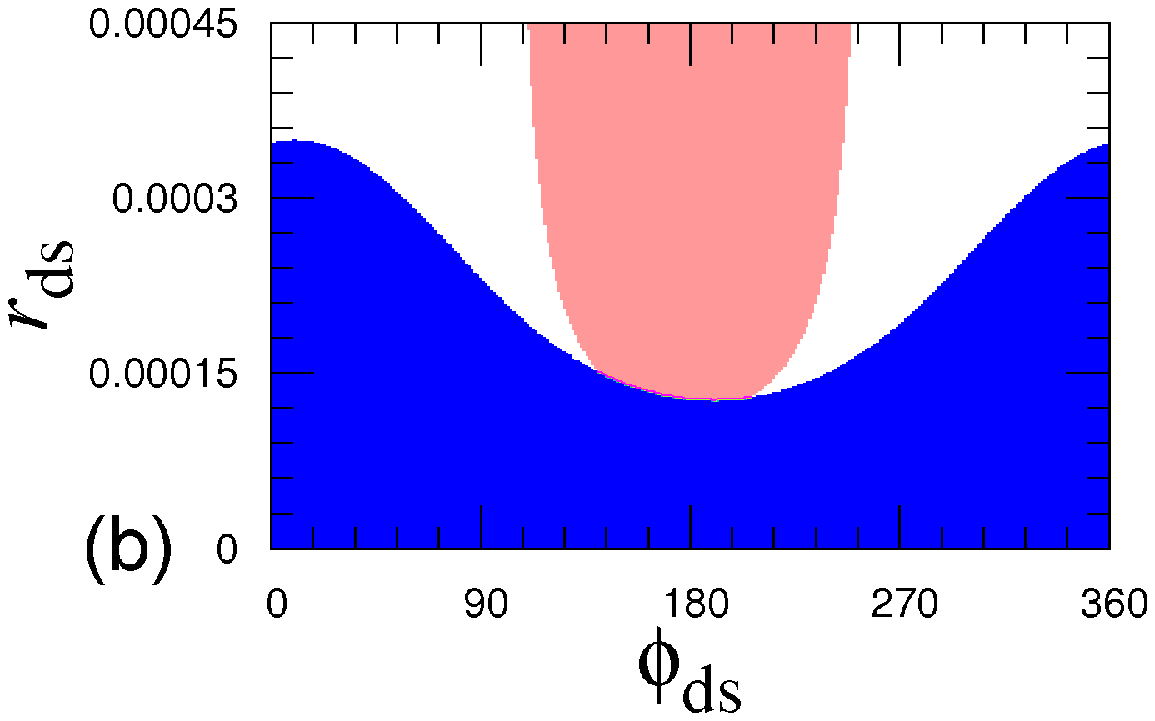}
\hspace{5mm}
\includegraphics[width=2.0in,angle=0]{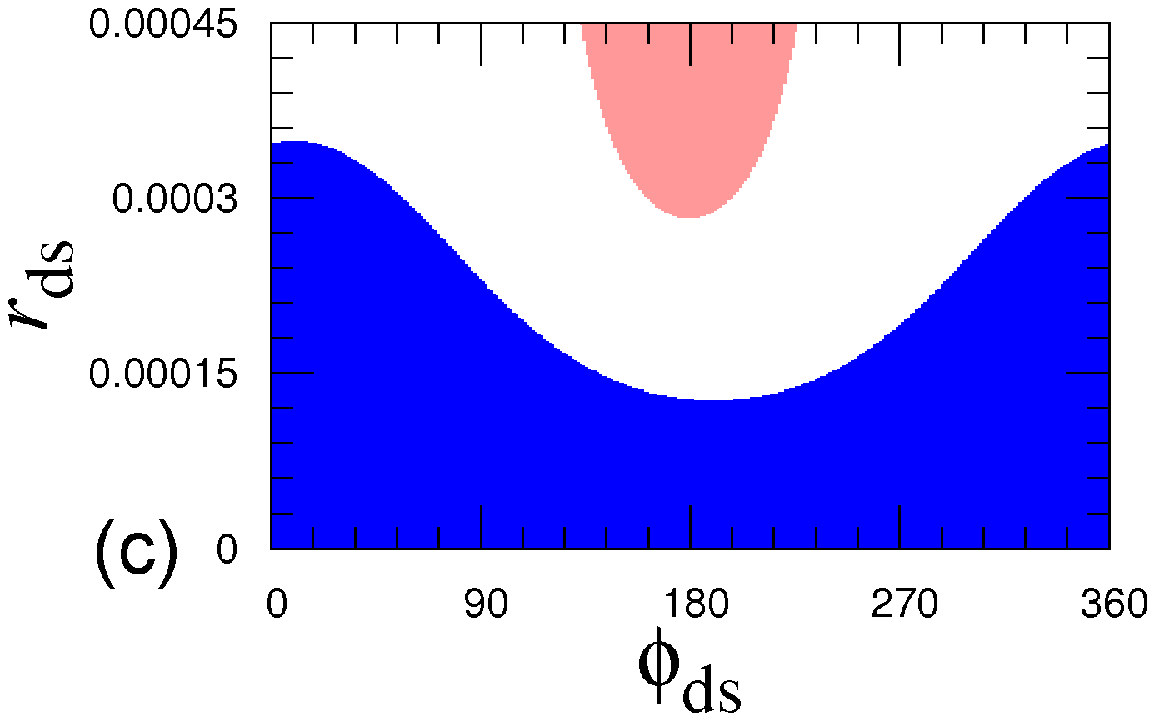}
 \caption{
 Allowed regions in $\phi_{ds}$-$r_{ds}$,
 where $V_{t'd}^{*}V_{t's} \equiv r_{ds}e^{i\phi_{ds}}$,
 for ${\cal B}(K^+\to\pi^+\nu\bar\nu)$ (blue or dark) and
 $D^0$--$\bar D^0$ mixing (pink or grey),
 for $m_{t'}=500$ GeV, $m_{b'}=480$ GeV,
  $V^*_{t's}V_{t'b} = 0.006\,e^{i75^\circ}$, and $|V_{t'b}|=$
 (a) 0.065, (b) 0.060, (c) 0.058. For the last
 $|V_{t'b}|$ value, the allowed regions no longer overlap,
 resulting in a lower bound close to 0.06.
}
\end{center}
\end{figure}

As illustrated in Fig.~2, 
we find that when $|V_{t'b}|$ drops below $0.06\;(0.12)$ for
$m_{t'}=500\;(300)$ GeV, the regions allowed by ${\cal
B}(K^+\to\pi^+\nu\bar\nu)$ and $\D m_D$ do not intersect anymore.
We conclude that, for $f_{B_s}\hat B_{B_s}^{1/2} = 266$ MeV,
%
\be
 |V_{t'b}| &\ge& 0.12, \ \ \ (m_{t'},\, m_{b'}) = (300,\, 280) \;{\rm GeV},
\ee 
for $V^*_{t's}V_{t'b} = 0.015\,e^{i81^\circ}$ (see Table I), and
%
\be
 |V_{t'b}| &\ge& 0.06, \ \ \ (m_{t'},\, m_{b'}) = (500,\, 480) \;{\rm GeV},
\ee 
for $V^*_{t's}V_{t'b} = 0.006\,e^{i75^\circ}$, where these are
meant as points of illustration only.

\begin{table}[t!]
\begin{center}
\begin{tabular}{ccc}
  \hline\hline
  nominal $V_{t'q}$ & $\ f_{B_s}\hat B_{B_s}^{1/2} = 266$ MeV
                    & $\ f_{B_s}\hat B_{B_s}^{1/2} = 295$ MeV \\
  \hline
  $m_{t'}=300$ GeV & $V_{t'b}=-0.18$ \ \ \ &  $V_{t'b}=-0.22$ \ \ \ \\
                   & $V_{t's}=-0.083\,e^{-i81^\circ}$ & $V_{t's}=-0.1136\,e^{-i70^\circ}$ \\
  \hline
  $m_{t'}=500$ GeV & $V_{t'b}=-0.10$ \ \ \ &  $V_{t'b}=-0.12$ \ \ \ \\
                   & $V_{t's}=-0.06\,e^{-i75^\circ}$ & $V_{t's}=-0.083\,e^{-i61^\circ}$ \\
  \hline\hline
\end{tabular}
 \caption{Nominal (and EWPT allowed) $V_{t'b}$ and $V_{t's}$ values
 for different $m_{t'}$ (with $m_{b'}$ taken 20 GeV lower)
 and $f_{B_s}\hat B_{B_s}^{1/2}$.}
\end{center}
\end{table}

For the $f_{B_s}\hat B_{B_s}^{1/2} = 295$ MeV case,
$|V_{t's}^*V_{t'b}|$ is much larger than the 266 MeV case
(Table I), which aggravates the above tension. Taking the central
values for $V^*_{t's}V_{t'b}$ from Table I, we summarize the lower
bounds for $|V_{t'b}|$ in Table II. Note that the 295 MeV case has
a much narrower range for $|V_{t'b}|$.

For illustration, we take the mean values for $|V_{t'b}|$ from
Table~II, combine again with the central values of
$V_{t's}^*V_{t'b}$ from Table I, and give some ``nominal" values
for $V_{t'b}$ and $V_{t's}$, within the parametrization of the
$4\times 4$ CKM matrix of Ref.~\cite{HSS87} in Table~III. In the
Appendix, we show that $|V_{t'b}|$ values near the bounds of
Table~II are less favored by EWPT. Furthermore, larger
$|V_{t'b}|,\; |V_{t's}|$ imply larger $\chi^2$. Thus, the lower
$f_{B_s}\hat B_{B_s}^{1/2} \sim 266$ MeV case is probably more
welcome.

\section{\boldmath $K_L\to\pi^0\nu\bar\nu$ and $\sin2\Phi_D$}

In Ref.~\cite{HNS05}, $\ve'/\ve$ was utilized as a constraint, and
nonstandard hadronic parameter solutions were found for $m_{t'}
\sim 300$ GeV. But as we allow $m_{t'}$ to vary, it becomes
apparent that huge hadronic uncertainties preclude the utility of
$\ve'/\ve$ in providing a constraint. Instead, it may be more
interesting to illustrate the potential impact of a future
measurement of $K_L \to \pi^0\n\bar\n$, which is dominated
purely by short distance. The SM predicts ${\cal
B}^{\rm SM}(K_L \to \pi^0\nu\bar\nu) = (2.8 \pm 0.4) \times
10^{-11}$~\cite{Mescia}, while the current limit is ${\cal B}^{\rm
exp}(K_L \to \pi^0\nu\bar\nu) < 6.7 \times 10^{-8}$~\cite{klpinn}.
The E14 (now KOTO) experiment, however, proposes to conduct a
three-year physics run beginning in 2011, to reach of order 10
events if SM holds. Suppose 100-250 events are observed
(which would be spectacular), it would
imply ${\cal B}^{\rm exp}(K_L \to \pi^0\nu\bar\nu) \sim 1 \times 10^{-9}$.
This value is just below the Grossman--Nir
bound~\cite{GNbound}, $\ie$ ${\cal B}(K_L \to \pi^0\n\bar\n)/{\cal
B}(K^+ \to \pi^+\n\bar\n) \sim 4.4$, assuming that ${\cal
B}(K^+ \to \pi^+\n\bar\n)$ is itself on the higher side of
the current experimental central value.

Let us take $m_{t'} = 500$ GeV and $f_{B_s}\hat B_{B_s}^{1/2} =
266$ MeV for illustration. We plot in Fig.~3(a) the allowed
regions for ${\cal B}^{\rm exp}(K_L \to \pi^0\n\bar\n) \sim 1
\times 10^{-9}$ and $\ve_K^{\rm exp} = (2.229 \pm 0.012) \times
10^{-3}$~\cite{PDG}, with ${\cal B}(K^+ \to \pi^+\nu\bar\nu)$ as
the broad backdrop (it can be viewed as interfaced with
$D^0$--$\bar D^0$ mixing, e.g. Fig.~2(a)). Again
$V_{t'd}^{*}V_{t's} \equiv r_{ds}e^{i\phi_{ds}}$. We find two
possible solutions of $V_{t'd}$. However, one solution is ruled
out by the constraint $\sin2\Phi_{B_d}^{\rm exp} = 0.672 \pm
0.023$~\cite{HFAG07} (see Fig.~3(c); note that with $\phi_{ub}
\simeq \phi_3 \cong 60^\circ$, $\sin2\Phi_{B_d}^{\rm SM} \simeq
0.687$ is expected), where an improvement of error by factor of 3
is also illustrated. Comparing Fig.~3(a) and Fig.~3(c), the only
possible solution is $V_{t'd} \sim -0.0032\,e^{-i18^\circ}$. {\it
This would in fact complete the $4\times 4$ CKM matrix}.

As further corollary to the full determination of the $4\times 4$
CKM matrix, let us see how the value for $\sin2\Phi_D$ is
correlated with $K_L \to\pi^0\nu\bar\nu$. For this purpose, we
parameterize 
$M_{12}^{D}$ as
\be
 M_{12}^{D} &=&
  \frac{G_F^2M_W^2}{12\pi^2}\, m_D f_D^2 B_D\,\eta(m_c,M_W) \no\\
 &\,&\times\,(\lambda_{b'}^2+R_{LD})S_0(x_{b'}),
 \label{m12d0_part2}
\ee
and for simplicity, we assume $R_{LD}$ to be real (this may not be
a very good assumption because the second type of LD effect in
Eq.~(\ref{m12d0}) could involve $\lambda_{b'}$ linearly).
This allows, by varying within $m_{b'} = 460 \pm 30$ GeV, to find
$\sin2\Phi_D \simeq 0.13$, and $|\cos2\Phi_D| \simeq 0.99$, which
are consistent with current data~\cite{HFAG07}. These values can
serve as a corollary for consistency check in the future. But it
should be clear that one would need to find a better handle on LD
effects.

\begin{figure}[t!]
\begin{center}
\includegraphics[width=2.0in,angle=0]{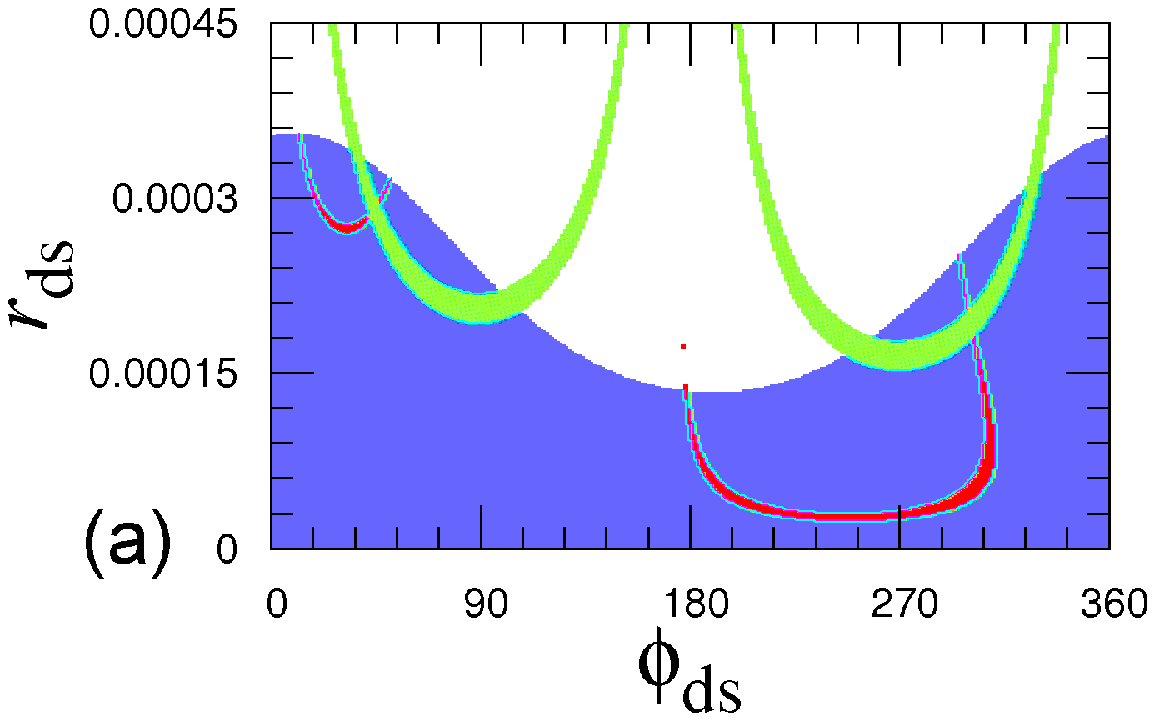}
\hspace{5mm}
\includegraphics[width=2.0in,angle=0]{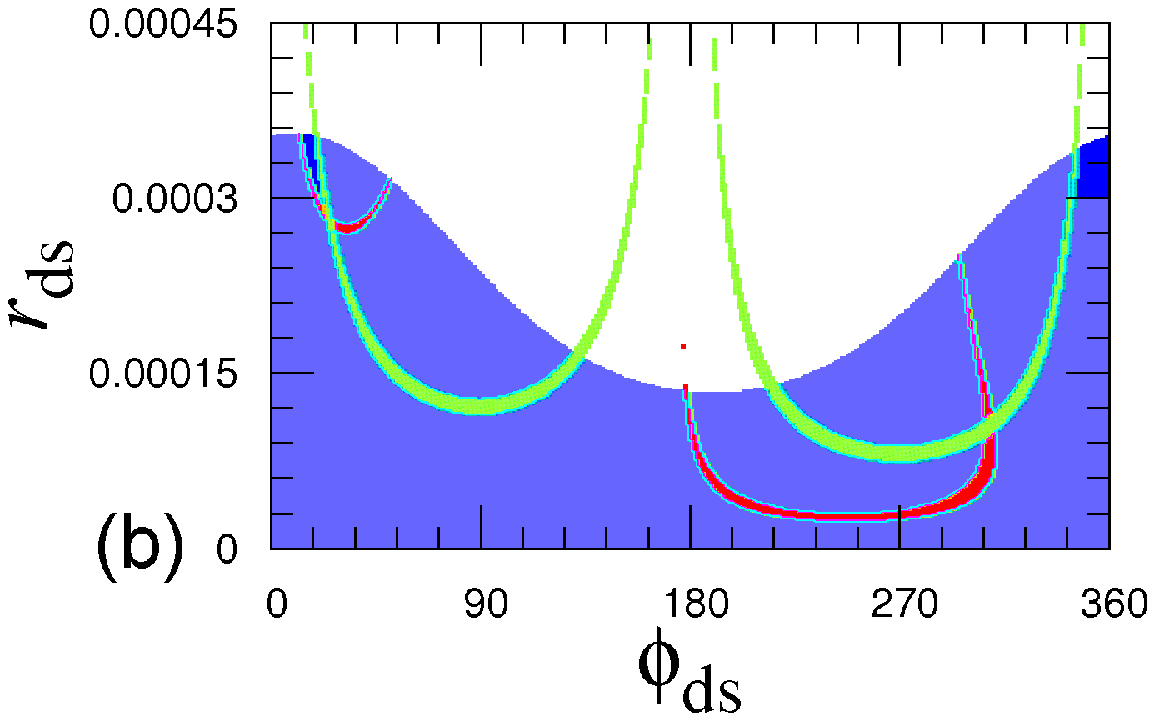}
\hspace{5mm}
\includegraphics[width=2.0in,angle=0]{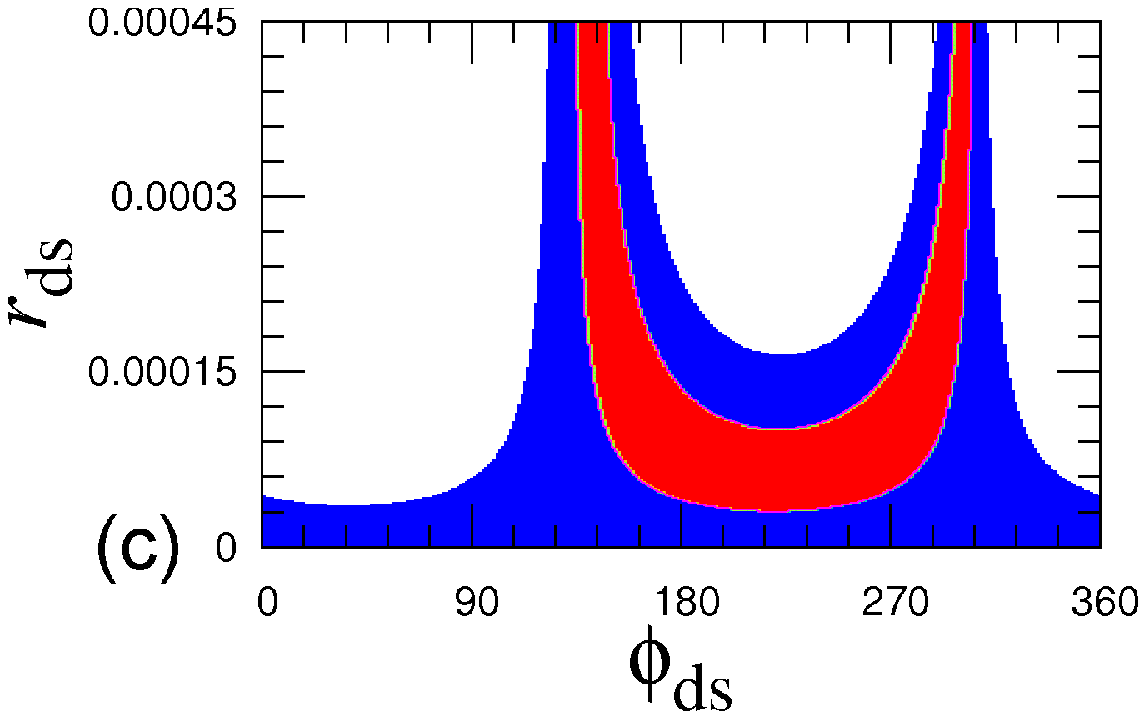}
 \caption{
  Impact of future measurements of (green or light)
  ${\cal B}(K_L\to\pi^0\nu\bar\nu) \sim$
    (a) $1 \times 10^{-9}$ and
    (b) $3 \times 10^{-10}$,
  for $m_{t'} = 500$ GeV, $m_{b'} = 480$ GeV, and
  $V_{t'b}=-0.10$, $V_{t's} = -0.060\,e^{-i75^\circ}$,
  together with $\varepsilon_K$ (red and dark grey), and
  combined ${\cal B}(K^+ \to \pi^+\nu\bar\nu)$ and
   $D^0$--$\bar D^0$ mixing (blue or dark).
  The allowed range for $\sin2\Phi_{B_d}^{\rm exp} = 0.672 \pm 0.023$
  (blue or dark) is given in (c),
  together with a more precise $0.672 \pm 0.008$ (red or dark grey).
}
\end{center}
\end{figure}

To illustrate a smaller value for ${\cal B}^{\rm exp} (K_L \to
\pi^0\n\bar\n)$, we take the value of $3 \times 10^{-10}$ (still
10 times the SM value) and replot in Fig.~3(b). Compared with
3(a), it can be noted that the two branches for ${\cal B}^{\rm
exp}(K_L \to \pi^0\n\bar\n)$ are less symmetric and each less
parabolic. This is simply because for Fig.~3(a), the 4th
generation effect is predominant, hence the allowed lowest
$r_{ds}$ value is for $\phi_{ds}$ purely imaginary. For the lower
${\cal B}^{\rm exp}(K_L \to \pi^0\n\bar\n)$ case of Fig.~3(b), the
top effect matters more, causing some qualitative change.
In any case, for the intersection of the allowed regions of ${\cal
B}^{\rm exp}(K_L \to \pi^0\n\bar\n) \sim 3\times 10^{-10}$ and
$\ve_K$ in Fig.~3(b), we find $V_{t'd} \sim
-0.0018\,e^{-i22^\circ}$, and a much smaller imaginary part for
$\lambda_{b'} = V_{ub'}^*V_{cb'}$, hence $\sin2\Phi_D$ would drop
considerably. This can be understood by noting that $V_{ub'}$ is
now dominated by the second term in Eq.~(\ref{vubp}), i.e.
$|s_{14}/s_{12}s_{24}| \sim 0.14$, while $V_{cb'}$ is always
dominated by the first $s_{24}$ term in Eq.~(\ref{vcbp}), hence
the large associated phase of $\phi_{sb}$ largely cancels. The
long distance $R_{LD} $ effect would only further dilute
$\sin2\Phi_D$. Therefore, we do not quote any value for
$\sin2\Phi_D$, except that, if ${\cal B}^{\rm exp}(K_L \to
\pi^0\n\bar\n)$ is on the low side, then one should expect
$\sin2\Phi_D$ to be rather small as well.
Note that, as can be seen from Fig. 3(c), if the central value for
$\sin2\Phi_{B_d}$ remains, but with error reduced by a factor of
3, tension would arise. Thus, future $\sin2\Phi_{B_d}$ measurement
would provide a crosscheck.

\begin{table}[t!]
\begin{center}
\begin{tabular}{cc}
  \hline\hline
  Real $R_{LD}$  &  \\
  ${\cal B}_{K_L\to\pi^0\n\bar\n} = 9.2 \times 10^{-10}$
   & $V_{t'd} = -0.0032\,e^{-i18^\circ}$ \\
  $\ve_K = 2.229 \times 10^{-3}$
   & ${\cal B}_{K^+ \to \pi^+\n\bar\n} = 2.1 \times 10^{-10}$ \\
  \hline
  $x_D = 9.1 \times 10^{-3}$ &  \\
  $m_{t'} = 500$ GeV
   &  $|\l_{b'}^2 + R_{LD}| = (16.2^{-1.6}_{+1.9}) \times 10^{-7}$  \\
  $m_{b'} = 460 \pm 30$ GeV & $\l_{b'}^2 = (8.0 + 2.1i)\times10^{-7}$ \\
  $V_{t'b} = -0.10$ & $\sin2\Phi_D \simeq 0.13$\\
  $V_{t's} = -0.06\,e^{-i75^\circ}$
   & $|\cos2\Phi_D| \simeq 0.99$\\
  \hline\hline
\end{tabular}
 \caption{A scenario for future measurement of large
 ${\cal B}(K_L\to\pi^0\n\bar\n)$, where $\lambda_{b'} = V_{ub'}^*V_{cb'}$.
 Taking $R_{LD}$ as real, we can get
 $\sin2\Phi_{D}$ and $\cos2\Phi_D$ once a full $4\times4$ CKM matrix
 is determined, where we illustrate with a finite range for $m_{b'}$.
 The left-hand side are inputs.}
\end{center}
\end{table}

We summarize the more spectacular scenario of ${\cal B}^{\rm
exp}(K_L \to \pi^0\n\bar\n) \sim 1 \times 10^{-9}$, which nearly
saturates the Grossman--Nir bound, in Table~IV. The more specific
values given in this Table are recalculated from the intersection
values for $\phi_{ds}$ and $r_{ds}$. We also give the approximate
values of the $4\times 4$ CKM matrix,
\be
 \left[
  \begin{array}{cccc}
    0.974 & 0.225 & 0.0036e^{-i60^\circ} & 0.015\,e^{i64^\circ} \\
    -0.226 & 0.972 & 0.041 & 0.060\,e^{i72^\circ} \\
    0.008\,e^{-i22^\circ} & -0.043e^{-i7^\circ} & 0.994 & 0.099e^{-i1^\circ} \\
    -0.003e^{-i18^\circ} & -0.06e^{-i75^\circ} & -0.1 & 0.993 \nonumber
  \end{array}
 \right]
\ee
which we do not aim at any precision, just to illustrate the
$m_{t'} = 500$ GeV case, and compare with the numerical values
given 5 years ago in Ref.~\cite{HNS05} for the $m_{t'} = 300$ GeV
case. As discussed, this is for an optimal value for ${\cal B}(K_L
\to \pi^0\n\bar\n)$ for the future measurement at the KOTO
experiment. If the measured value for ${\cal B}(K_L \to
\pi^0\n\bar\n)$ is lower, then the strength and phase of $V_{t'd}$
would further drop, the details depending also on the intersection
with $\ve_K$ as well as the precise $m_{t'}$ value. But the
$V_{ub'}$ value would be less affected. Note also that ${\cal
B}(K^+ \to \pi^+\n\bar\n) = 2.1 \times 10^{-10}$ is a little on
the high side compared to current measurement, but not by too
much.
Of course, if measurement of $\sin2\Phi_D$ could get ahead of
${\cal B}^{\rm exp}(K_L \to \pi^0\n\bar\n)$, information of
$V_{t'd}$ can also be extracted. But it would depend on our
understanding of the LD effects, which appears difficult. From our
discussion, we also see that a larger $\sin2\Phi_D$ value would
likely imply a large ${\cal B}(K_L \to \pi^0\n\bar\n)$.

\section{Discussion and Conclusion}

A 4th generation is a very natural extension of the Standard
Model, as we already have 3 generations. It is curious why the
famed measurement of $\sin2\Phi_{B_d}$ at the B factories came out
consistent with SM3, while there is also the tension in EWPT
measurements. However, with the LHC finally starting, we are
entering an era where the question of whether there is a 4th
generation can be answered once and for all~\cite{AH2006} by
direct search. This paper surveys the flavor and CPV aspects,
focusing on where information may be extracted. For this reason,
we have not used the experimentally established $\Delta A_{K\pi}$,
nor $\ve'/\ve$, as these are marred by long-distance or hadronic
effects. We did use the $\Delta m_D$ measurement. Although LD
effects also enters, the measured strength still puts a constraint
on the combination of $|V_{ub'}^*V_{cb'}|m_{b'}^2$.

We illustrated with a series of steps on how a full $4\times 4$
CKM matrix can be determined, from the present towards the future.
We took mainly $m_{t'}=500$ GeV, $m_{b'}=480$ GeV and $f_{B_s}\hat
B_{B_s}^{1/2} = 266$ MeV as an example.
First, combining the constraints of $\D m_{B_s}$ and ${\cal B}(b
\to s\ell\ell)$, where the nondecoupling nature of the $t'$ quark
could make its effect felt, one could determine $V_{t's}^*V_{t'b}
\sim 0.006\,e^{i75^\circ}$. This leads to a predicted range for
$\sin2\Phi_{B_s}$, the measurement of which is of great current
interest at the Tevatron and LHC. In turn, once $\sin2\Phi_{B_s}$
is measured with suitable precision, it would provide us with a
probe of $V_{t's}^*V_{t'b}$, although $\D m_{B_s}$ would still be
marred by $f_{B_s}$, and we would still rely on measurements such
as ${\cal B}(b \to s\ell\ell)$.
Second, $R_b$ gives rise to an upper bound of $|V_{t'b}|<0.13$,
and from combining ${\cal B}(K^+ \to \pi^+\n\bar\n)$ and
$D^0$--$\bar D^0$ mixing, one could extract a lower bound of
$|V_{t'b}| > 0.06$. This follows from the assumption that
$|V_{t's}^*V_{t'b}|$ is known. Then, a lower $|V_{t'b}|$ means a
higher $|V_{t's}|$. The bound from ${\cal B}(K^+ \to
\pi^+\n\bar\n)$ then demands a smaller $|V_{t'd}|$, which in turn
limits the ability for $|V_{ub'}|$ to satisfy the $\Delta m_D$
constraint. In the Appendix, we show that the bounds on
$|V_{t'b}|$ is consistent with EWPT constraints, but the central
value is (and generally, smaller $|V_{t'b}|$ and $|V_{t's}|$
values are) preferred. For sake of illustration, we offer $V_{t'b}
= -0.10$ and $V_{t's} = -0.06\,e^{-i75^\circ}$ (in the
parametrization of Ref.~\cite{HSS87}) as nominal values for
$m_{t'}$, $m_{b'} =500$, 480 GeV.

There is insufficient information at present to pin down
$V_{t'd}$, but this can be achieved with a future measurement of
$K_L\to\pi^0\n\bar\n$. Suppose ${\cal B}(K_L\to\pi^0\n\bar\n) =
10^{-9}$ is found by the KOTO experiment. With the current data on
${\cal B}(K^+ \to \pi^+\n\bar\n)$, this is close to saturating the
Grossman--Nir bound, so it is probably optimistic. By combining
with $\ve_K$ as a constraint, we get two possible solutions of
$V_{t'd}$. Then, taking into account the constraint of
$\sin2\Phi_{B_d}$ (the measurement of which should also improve),
this selects out the solution $V_{t'd}=-0.0032\,e^{-i18^\circ}$
(again in the parametrization of Ref.~\cite{HSS87}). So, it seems
that within a decade, we may determine the complete $4\times4$ CKM
matrix.

For the time being, with LHC experiments soon to catch up with the
vigorous pursuit of the measurement of $\sin2\Phi_{B_s}$ and
direct $t'$, $b'$ search at the Tevatron, if we consider the
uncertainties from $f_{B_s}\hat B_{B_s}^{1/2}$ and ${\cal B}(b\to
s\ell\ell)$, $\sin2\Phi_{B_s}$ can range from $-0.4$ to $0$. As
the $t'$ mass bound rises, one expects a weaker, but still
negative, $\sin2\Phi_{B_s}$. We see that the critical future
measurement beyond $\sin2\Phi_{B_s}$ would be ${\cal
B}(K_L\to\pi^0\n\bar\n)$, which is also purely short distance, and
can help us determine $V_{t'd}$. The measurement of
$\sin2\Phi_{B_d}$ by all means should also be improved. The usage
of CPV in $D$ mixing, $\sin2\Phi_D$, would require knowledge of
long distance effects.

\vskip 0.3cm \noindent{\bf Note Added}.\ While writing this paper,
similar discussions have also been made by Soni {\it et
al.}~\cite{Soni10} and Buras {\it et al.}~\cite{Buras10}, with
differences in emphasis than our approach.

\vskip 0.3cm \noindent{\bf Acknowledgement}.\ The work of WSH is
supported in part by NSC97-2112-M-002-004-MY3 and NTU-98R0066. The
work of CYM is supported by NSC98-2811-M-002-103.

\section*{Appendix}

In a recent paper by Chanowitz~\cite{Chanowitz2}, the 4th
generation  corrections to the oblique parameters $S,\;T$ were
considered, which enter
\be
 && M_W^2 = \left(M_W^{\rm SM}\right)^2\,
         \left[1 - \frac{\alpha\Delta S}{2(c_w^2-s_w^2)}
                 + \frac{c_w^2\alpha \Delta T}{(c_w^2-s_w^2)}\right],
 \label{mw2cor} \\
 && \SWTWOQFBP =
    \SWTWOQFBP\vert^\SM\, \no\\
 &&\hspace{1.8cm} \left[1 + \frac{\alpha\Delta S}{4s_w^2(c_w^2-s_w^2)}
                          - \frac{c_w^2\alpha \Delta T}{(c_w^2-s_w^2)}\right],
 \label{sw2effcor}
\ee
and
\be
 \Gamma(Z\to\nu\bar\nu) = \Gamma^{\rm SM}(Z\to\nu\bar\nu)\,
                          \left[1 + \alpha \Delta T\right].
 \label{Znn}
\ee
These formulas can be found in Ref.~\cite{Maksymick}. Here, we
neglect all other parameters $U,\;V,\;W,\;X,\;Y$, but we extend
formulas $\SWTWOQFBP$ to $\sin^2\theta^f_{\rm eff}$ and
$\Gamma(Z\to\nu\bar\nu)$ to $\Gamma(Z\to f\bar f)$. Though it is
not our main concern, following Chanowitz, we also wish to
investigate the impact of considering quark mixing on the
electroweak observables.

\begin{figure}[t!]
\begin{center}
\includegraphics[width=2.4in,angle=0]{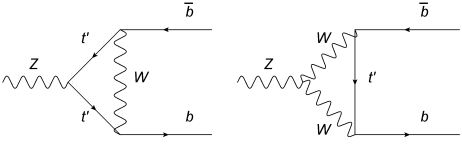}
 \caption{
 One-loop correction to $Zb\bar b$ vertex from $t'$.
 }
\end{center}
\end{figure}

Compared to $S,\;T$, which come from vacuum polarization
corrections, one also has to include the vertex corrections from
Fig.~4 (touched upon in Sec.~III for upper bound on $V_{t'b}$;
note that the vacuum polarization effects largely cancel in the
ratio of $R_b$), which cause the shift to $Zb\bar b$ couplings
\be
 v_b &=& v_b^{\rm SM} + \delta g_{bL}, \\
 a_b &=& a_b^{\rm SM} + \delta g_{bL},
\ee
where $v_b^{\rm SM} = -\frac{1}{2} + \frac{2}{3}s_w^2$, $a_b^{\rm
SM} = -\frac{1}{2}$, and $\delta g_{bL}$ is given in
Ref.~\cite{Chanowitz2}. Hence, the effective couplings
$g_V^b,\;g_A^b$ become
\be
 g_V^b &=& \sqrt{\rho_Z^b} \left(-\frac{1}{2}
                               + \frac{2}{3}\sin^2\theta^b_{\rm eff}\right)
         \frac{v_b}{v_b^{\rm SM}} \\
 g_A^b &=& \sqrt{\rho_Z^b} \left(-\frac{1}{2}\right)\frac{a_b}{a_b^{\rm SM}}.
\ee
Inserting this into the formula~\cite{Bernabeu} for $\Gamma(Z\to
q\bar q)$ is
\be
 \Gamma(Z\to q\bar q) = \frac{\alpha\,M_Z}{4s^2_wc_w^2}
                        \left(\vert a_q^{\rm SM}\vert^2 + \vert v_q^{\rm SM}\vert^2\right)
                        \left(1 + \delta_q^{(0)}\right) \cdots,
                        \no
\ee
we then have the complete correction formula
\be
 && \Gamma(Z\to b\bar b) = \Gamma^{\rm SM}(Z\to b\bar b)
                           \left[1 + \alpha \D T\right] \no\\
 &&\hspace{2.5cm}          \frac{|a_b|^2+|v_b|^2}{|a_b^{\rm SM}|^2+|v_b^{\rm SM}|^2}.
\ee

\begin{figure}[t!]
\begin{center}
\includegraphics[width=2.0in,angle=0]{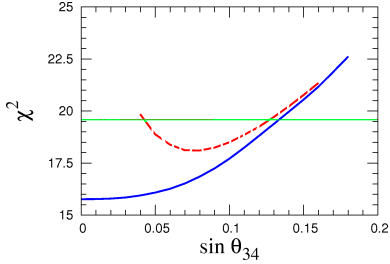}
 \caption{
 For $m_{t'}=500$ GeV, $m_{b'}=480$ GeV, $m_{\ell_4}=145$ GeV, $m_{\nu_4}=100$ GeV:
 solid (blue) $\chi^2$ vs $|V_{t'b}| = s_{34}$ for $s_{24}=s_{14}=0$; and
 dashed (red) $\chi^2$ vs $s_{34}$ for $V_{t's}^*V_{t'b}=0.006\,e^{i75^\circ}$.
 }
\end{center}
\end{figure}

We then follow the procedures given in Ref.~\cite{Chanowitz2}.
Neglecting $\Gamma_W$ fit and including the correlation matrices
in Ref.~\cite{EWWGReport}, we use ZFITTER 6.4.2 to successfully
reproduce the results of Ref.~\cite{Chanowitz2}.
With this attained, we take $m_{t'}=500$ GeV, $m_{b'}=480$ GeV,
$m_{\ell_4}=145$ GeV, $m_{\nu_4}=100$ GeV, and set
$s_{14}=s_{24}=0$, The plot of $\chi^2$ vs $s_{34}$ is given in
Fig.~5 as the solid (blue) curve. The best fit occurs at
$s_{34}=0$ with $\chi^2_{\rm min}=15.8$, and $95\%$ CL is located
at $\chi^2=19.6$.

Next, we consider the case of taking
$V_{t's}^*V_{t'b}=0.006\,e^{i75^\circ}$ (see Table I), as
motivated by our flavor and CPV analysis. We see from the dashed
(red) curve in Fig.~5 that $95\%$ CL is located at $s_{34}=0.04$
and $0.13$, with the lowest $\chi^2$ at $s_{34}=0.08$ (the lower
$s_{34}$ value would could trouble through a rather large
$V_{t's}$). The rise in $\chi^2$ away from $s_{34}=0.08$ is in
part due to fixing $|V_{t's}^*V_{t'b}|$ at 0.006. But with this
treated as external to the fit, the change in $\chi^2$ is not much
worse than treating the effect of $V_{t'b}$ in the loop but
ignoring $V_{t's}$. Note that the latter affects $Z\to s\bar s$,
but this process is hard to separate experimentally.


\begin{thebibliography}{99}
%
\bibitem{HHHMSU}
  B.~Holdom, W.-S.~Hou, T.~Hurth, M.L.~Mangano, S.~Sultansoy and G.~\"Unel,
  PMC Phys.\  A {\bf 3}, 4 (2009).
%
\bibitem{CDFDzero}
  See http://tevbwg.fnal.gov/results/Summer2009\_betas/;
  and as reported by G. Punzi at Europhysics Conference on
  High Energy Physics, Krakow, Poland, July, 2009
  [arXiv:1001.4886 [hep-ex]].
%
\bibitem{DAKpi}
  S.-W. Lin, Y. Unno, W.-S. Hou, P. Chang {\it et al.} [Belle collaboration],
  Nature {\bf 452}, 332 (2008).
%
\bibitem{PDG}
   C. Amsler {\it et al.} [Particle Data Group],
   Phys. Lett. B {\bf 667}, 1 (2008).
%
\bibitem{HLMN}
  W.-S.~Hou, M.~Nagashima and A.~Soddu,
  Phys.\ Rev.\ Lett.\  {\bf 95}, 141601 (2005);
  W.-S.~Hou, H.-n.~Li, S.~Mishima and M.~Nagashima,
  {\it ibid.}  {\bf 98}, 131801 (2007).
%
\bibitem{HNS07}
  W.-S.~Hou, M.~Nagashima, A.~Soddu,
  Phys.\ Rev.\ D {\bf 76}, 016004 (2007).
%
\bibitem{HNS05}
  W.-S.~Hou, M.~Nagashima, A.~Soddu,
  Phys.\ Rev.\ D  {\bf 72}, 115007 (2005).
%
\bibitem{He}
  H.-J. He, N. Polonsky and S.-f. Su,
  Phys. Rev. D {\bf 64}, 053004 (2001).
%
\bibitem{Kribs}
  G.D. Kribs, T. Plehn, M.S. Spannowsky and T.M.P. Tait,
  Phys. Rev. D {\bf 76}, 075016 (2007).
%
\bibitem{Chanowitz2}
  M.~Chanowitz,
  Phys. Rev. D {\bf 79}, 113008 (2009).
%
\bibitem{Erler}
  J.~Erler and P.~Langacker,
  arXiv:1003.3211 [hep-ph].
%
\bibitem{HouCJP}
  W.-S.~Hou,
  Chin.\ J.\ Phys.\ {\bf 47}, 134 (2009)
  [arXiv:0803.1234 [hep-ph]].
%
\bibitem{CDFtp}
  T.~Aaltonen {\it et al.}  [CDF Collaboration],
  Phys.\ Rev.\ Lett.\  {\bf 100}, 161803 (2008).
  Updates available on CDF webpage http://www-cdf.fnal.gov/.
%
\bibitem{CDFbp}
  T.~Aaltonen {\it et al.}  [CDF Collaboration],
  Phys.\ Rev.\ Lett.\  {\bf 104}, 091801 (2010).
%
\bibitem{HWS}
  W.-S.~Hou, R.S.~Willey and A.~Soni,
  Phys.\ Rev.\ Lett.\  {\bf 58}, 1608 (1987)
  [Erratum-ibid.\  {\bf 60}, 2337 (1988)].
%
\bibitem{HSS87}
  W.-S.~Hou, A.~Soni and H.~Steger,
  Phys. Lett. B {\bf 192}, 441 (1987).
%
\bibitem{latHPQCD09}
  E. Gamiz {\it et al.} [HPQCD Collaboration],
  Phys. Rev. D {\bf 80}, 014503 (2009).
%
\bibitem{Bobeth00}
  C.~Bobeth, M.~Misiak and J.~Urban,
  Nucl.\ Phys.\ B {\bf 574}, 291 (2000).
  Beware of a few typos (cf. arXiv:hep-ph/9910220).
%
\bibitem{Hiller}
  One could use the measurement of foward-backward asymmetry in $B \to K^* \ell\ell$,
  especially the large $q^2 \equiv m^2_{\ell\ell}$ part, to further constrain the 4th generation
  parameters; see C.~Bobeth, G. Hiller, and G. Piranishvili,
  JHEP {\bf 0807}, 106 (2008). This would be an interesting
  direction to pursue in the future, when the lower $q^2$
  discrepancy is better established and understood.
  For this aspect, see also A.~Hovhannisyan, W.-S.~Hou and N.~Mahajan,
  Phys.\ Rev.\  D {\bf 77}, 014016 (2008).
%
\bibitem{Yanir}
  T.~Yanir, JHEP{\bf 0206}, 044 (2002).
%
\bibitem{BSU}
  A.J.~Buras, F.~Schwab and S.~Uhlig,
  Rev.\ Mod.\ Phys.\ {\bf 80} 965 (2008).
%
\bibitem{Artamonov}
  V.A.~Artamonov {\it et al.} [E949 Collaboration],
  Phys.\ Rev.\ Lett.\  {\bf 101}, 191802 (2008).
%
\bibitem{Bobrowski}
  M.~Bobrowski, A.~Lenz, J.~Riedl and J.~Rohrwild,
  Phys.\ Rev.\ D {\bf 79}, 113006 (2009).
%
\bibitem{Petrov}
  Besides considerations in Refs.~\cite{HNS05} and \cite{HNS07},
  a standard earlier reference is
  E. Golowich, J. Hewett, S. Pakvasa, A.A. Petrov,
  Phys.\ Rev.\ D {\bf 76}, 095009 (2007).
%
\bibitem{HFAG08}
  E.~Barberio {\it et al.}  [Heavy Flavor Averaging Group (HFAG)],
  arXiv:0808.1297 [hep-ex].
%
\bibitem{Mescia}
  F.~Mescia, C.~Smith,
  Phys.\ Rev.\ D {\bf 76}, 034017 (2007).
%
\bibitem{klpinn}
  J.~Ahn {\it et al.}  [E391a Collaboration],
  Phys.\ Rev.\ Lett.\ {\bf 100}, 201802 (2008);
%
\bibitem{GNbound}
  Y. Grossman and Y. Nir,
  Phys. Lett. B {\bf 398}, 163 (1997).
%
\bibitem{HFAG07}
  E.~Barberio {\it et al.}  [Heavy Flavor Averaging Group (HFAG)],
  arXiv:0704.3575 [hep-ex].
%
\bibitem{AH2006}
  A.~Arhrib and W.-S.~Hou,
  JHEP {\bf 0607}, 009 (2006).
%
\bibitem{Soni10}
  A.~Soni, A.K.~Alok, A.~Giri, R.~Mohanta and S.~Nandi,
  arXiv:1002.0595 [hep-ph].
%
\bibitem{Buras10}
  A.J.~Buras, B.~Duling, T.~Feldmann, T.~Heidsieck, C.~Promberger and S.~Recksiegel,
  arXiv:1002.2126 [hep-ph].
%
\bibitem{Maksymick}
  I.~Maksymyk, C.P.~Burgess and D. London,
  Phys.\ Rev.\ D  {\bf 50}, 529 (1994).
%
%
\bibitem{Bernabeu}
  J.~Bernab\'eu, A.~Pich and A. Santamaria,
  Nucl.\ Phys. {\bf 363}, 326 (1991).
%
%
\bibitem{EWWGReport}
  The ALEPH, DELPHI, L3, OPAL and SLD Collaborations,
  the LEP Electroweak Working Group, and
  the SLD Electroweak and Heavy Flavour Groups,
  Phys.\ Rept.\ {\bf 427}, 257 (2006).

%
\end{thebibliography}
\end{document}